\documentclass[showkeys,showpacs,superscriptaddress,nofootinbib,prd]{revtex4-2}
\usepackage{amsmath}
\usepackage{amssymb}
\usepackage{graphicx}
\usepackage{yfonts}
\usepackage{ulem,xcolor}
\usepackage{color,soul}
\usepackage{amsthm}

\newtheorem{theorem}{Theorem}[section]
\theoremstyle{definition}
\newtheorem{definition}{Definition}[section]

\begin{document}
	
\title{Quantization of measure in gravitation
	}
	
\author{
	Vladimir Dzhunushaliev
}
\email{v.dzhunushaliev@gmail.com}
	
\affiliation{
	Department of Theoretical and Nuclear Physics,  Al-Farabi Kazakh National University, Almaty 050040, Kazakhstan
}

\affiliation{
	Institute of Experimental and Theoretical Physics,  Al-Farabi Kazakh National University, Almaty 050040, Kazakhstan
}

\affiliation{
	Academician J.~Jeenbaev Institute of Physics of the NAS of the Kyrgyz Republic, 265 a, Chui Street, Bishkek 720071, Kyrgyzstan
}
	
\author{Vladimir Folomeev}
\email{vfolomeev@mail.ru}

\affiliation{
	Institute of Experimental and Theoretical Physics,  Al-Farabi Kazakh National University, Almaty 050040, Kazakhstan
}

\affiliation{
	Academician J.~Jeenbaev Institute of Physics of the NAS of the Kyrgyz Republic, 265 a, Chui Street, Bishkek 720071, Kyrgyzstan
}
	
\date{\today}
	
\begin{abstract}
Using the Radon-Nikodym theorem concerning the relation between any two measures, as well as the methods employed in loop quantum gravity, 
it is shown that, in gravitation, one can quantize any measure which is not even associated with metric. We have considered the simplest case 
where the proportionality coefficient between two operators of measure (the Radon-Nikodym derivative) is some function. 
The result is that in the right-hand side of the commutation relations for the measure the fundamental length becomes a variable quantity,
and this can lead to smoothing the singularity. The case where the Radon-Nikodym derivative is an operator is also under discussion. 
Classical and quantum theories on the background of space endowed with quantum measure are under consideration.
\end{abstract}
	
	
\keywords{measure, quantization, field theory, background
}
	
\date{\today}
	
\maketitle
	
\section{Introduction}
In differential geometry, there are only three geometric quantities which can be defined on a manifold~--
the measure, connection, and metric. In general relativity, the first two quantities are associated with metric:
the measure is defined by the square root of the determinant of a metric tensor and the connection is the Christoffel symbols.
But there is the following difference between the measure and connection. The connection associated with metric
arises naturally in general relativity when one uses the Palatini approach. By varying the Einstein-Hilbert action
with respect to the metric and connection, one gets equations for the connection which ensure that the connection becomes associated with metric.
 It is interesting that such a procedure for the measure is absent.
This means that: (a)~in general relativity, the measure  is merely postulated as $d \mu = \sqrt{-g} \, dx^0 dx^1 dx^2 dx^3$; and
(b)~for the measure, there are no any field equations, in contrast to the connection for which there are the corresponding equations
appearing when one uses the Palatini method. It is easily explained by the fact that between the measure on the one hand, and
the metric and connection on the other, there is the following difference: the measure is a nonlocal quantity, whereas the metric and connection
are local ones.

An entirely different situation arises in loop quantum gravity (LQG) where, when quantizing the Ashtekar variables,
there emerges the quantization of the volume, area, and length (see, e.g., the textbooks~\cite{Rovelli:2014ssa,Gambini:2011zz}), i.e.,
the quantization of the measure. This remarkable fact suggests that, apparently, in quantum gravity the measure becomes a
dynamical variable and is defined while quantising, although there are no any dynamical equations for the measure.

One of the fruitful approaches in quantum gravity is LQG~\cite{Rovelli:1987df,Rovelli:1989za}, where the spectra of such geometric quantities like the area~\cite{Rovelli:1994ge} and volume~\cite{Ashtekar:1996eg} have been calculated. In LQG, the SU(2) Ashtekar-Barbieri connection is introduced; this enables one to quantize the operators of area and volume. For a review of main results obtained in LQG, see Ref.~\cite{Ashtekar:2021kfp}.

The purpose of the present paper is to apply the methods developed in LQG for quantizing an arbitrary measure. 
We use the results obtained to investigate the dynamics of arbitrary classical and quantized fields on the background of space endowed with quantum measure.
In conclusion, we discuss questions arising in quantizing the measure. 

\section{Quantization of measure in LQG }

In the present paper we consider the question of the quantization of measure in gravitation. This implies that the measure of any region $\mathcal{A}$ in space, for example, of volume, area or length, becomes an operator:
$$	\widehat{\mu (\mathcal{A})} = \int \limits_{\mathcal{A}} \widehat{d \mu} ,$$
where $\mathcal{A}$ can be a volume, area or length. Of course, the main question here is how one can define an operator algebra  $\widehat{\mu (\mathcal{A})}$? One of the main problems here is that such an operator is nonlocal, since it is defined not at the point but in some region  $\mathcal{A}$. If we compare this situation with quantum field theory, we see that there fields are described by local operators and one knows how they can be quantized;  this is the difference of principle between quantum field theory and the problem considered in the present paper when we quantize a nonlocal operator of the measure
$\widehat{\mu (\mathcal{A})}$.

In quantum field theory, the question of quantization of the given theory is resolved using the fact that there is classical field theory with the
corresponding field equations, and this theory should be quantized by introducing field operators. To quantize the measure, such an approach is excluded,
since, in classical gravity, there are no field equations for the measure. In general relativity, the measure is merely postulated as
$$	d \mu_{GR} = \sqrt{-g} \, d\Omega ,$$
where $d\Omega = dx^0 dx^1 dx^2 dx^3$.

When quantizing the measure, there arise problems which are absent in local quantum field theory. For example, one has to take into account that in calculating the two-point Green function
$$	G\left(
		\widehat{\mu (\mathcal{A}_1)}, \widehat{\mu (\mathcal{A}_2)}
	\right)  =
	\left\langle
		Q \left| \widehat{\mu (\mathcal{A}_1)} \widehat{\mu (\mathcal{A}_2)} \right| Q
	\right\rangle$$
the regions $\mathcal{A}_1$ and $\mathcal{A}_2$ may intersect or not, they may have different dimensions, etc.
The essence of these problems is due to the fact that the measure is a nonlocal object, whereas in standard quantum field theory all physical quantities are local.

The remarkable fact is that the problems of quantization of such a nonlocal quantity as the measure had been solved in LQG.
But for this purpose, in LQG, some new extra structure is introduced, which is absent in classical general relativity~-- a triangulation $\Delta$ of our spacetime
 $\mathcal{M}$, i.e., the spacetime is divided into four-simplices. Every four-simplex is bounded by five tetrahedra. In LQG, the dual $\Delta^*$ of the triangulation with vertices, edges, and faces are used.
The relation between $\Delta$ and $\Delta^*$ is illustrated as follows:
\begin{equation}
	\begin{matrix}
	\text{Bulk triangulation } \Delta &  & \text{Dual complex}  \Delta^*\\
	\text{4-simplex } (v) &  \leftrightarrow& \text{vertex }  (\mathrm{v})\\
	\text{tetrahedron } (\tau) &  \leftrightarrow& \text{edge }  (\mathrm{e})\\
	\text{triangle } (t) &  \leftrightarrow& \text{face }  (\mathrm{f})\\
	\text{segment } (s) &  & \\
	\text{point } (p) &  &
	\end{matrix}\nonumber
\end{equation}
That is, instead of the triangulation $\Delta$, one can employ the graph $\Delta^*$ with vertices, edges, and faces.

This extra structure $\Delta^*$ enables one to resolve the problem of quantization in the following manner. One introduces the operator
(here, we follow the textbook~\cite{Rovelli:2014ssa}, chapter 7)
$$	\hat{\vec{E}}_{\ell} = 8 \pi \gamma G \hbar \hat{\vec{L}}_{\ell} ,$$
where $\gamma$ is the Barbero-Immirzi constant. The operator $\hat{\vec{L}}_{\ell}$ is the area operator of the faces of the tetrahedron punctured by the link $\ell$,
and, before the quantization (in the classical case), it is defined as follows:
$$	L^i_{S} = \frac{1}{2 \gamma} \int dS^i
	= \frac{1}{2 \gamma} \epsilon^i_{j k}
	\int \limits_{S} e^j \wedge e^k ,$$
with the tetrad $e^a_\mu$ chosen in the time gauge where $e^0 = dt$, $e^i = e^i_\mu dx^\mu$ and
$ i,j,k = 1,2,3$; $S$ is some two-dimensional surface punctured by the link $\ell$. The area of this surface is defined as
$$	S = \left| \vec{E}_{\ell} \right| .$$
After the quantization, the quantity $L^i_{\mathcal{S}}$ becomes an operator with the following commutation relations
that can be obtained from the hamiltonian analysis of general relativity, by promoting Poisson brackets to operators, or can be postulated~\cite{Barbieri:1997ks}:
\begin{equation}
	\left[
		\hat{E}^i_{S_1}, \hat{E}^j_{S_2}
	\right] = i \Delta_{S_1 S_2} l_0^2
	\epsilon^{i j}_{\phantom{i j} k} \hat{E}_{S_1}^k ,
\label{2_80}
\end{equation}
where $l_0^2 = 8 \pi \gamma G \hbar $ is a constant proportional $\hbar$ and with the dimensions of an area; $\Delta_{S_1 S_2}$ is the Dirac delta functional defined as
$$	\Delta_{S_1 S_2} =
	\begin{cases}
		1 \text{ if }	S_1 \text{ coincides with } S_2, \\
		0 \text{ otherwise }
	\end{cases}.$$
The eigenvalues of the square of this operator form a discrete spectrum
\begin{equation}
	S_j = l_0^2 \sqrt{j (j + 1)} .
\label{2_90}
\end{equation}
From the point of view of quantization of measure, the operator
$
	\frac{l_0^2}{2 \gamma} \epsilon^i_{j k} \hat{e}^j \wedge \hat{e}^k
$ is the operator of the measure $\widehat{d \mu^i_g}$ of the area spanned on the vectors $\hat{e}^j, \hat{e}^k$:
$$	\widehat{d \mu_g^i} = \frac{l_0^2}{2 \gamma} \epsilon^i_{j k} \hat{e}^j \wedge \hat{e}^k .$$
Here, we use the index $g$ in the definition of the element of the measure $\widehat{d \mu_g^i}$ in order to emphasize that this measure is associated with
the metric $g$. Using this definition of the operator  of the element of the measure, one can rewrite the commutation relations~\eqref{2_80} in the infinitesimal form,
\begin{equation}
	\left[
	\widehat{d \mu_g^i}, \widehat{d \mu_g^j}
	\right] = i \Delta_{dS_1 dS_2} l_0^2
	\epsilon^{i j}_{\phantom{i j} k} \widehat{d \mu_g^k} .
\label{2_110}
\end{equation}
Using this notion of the measure $\mu$, one can also rewrite the commutation relations~\eqref{2_80} in the following form:
\begin{equation}
	\left[
	\widehat{\mu_g}\left( {S^i_1}\right) , \widehat{\mu_g}\left( {S^j_2}\right)
	\right] = i \Delta_{S_1 S_2} l_0^2
	\epsilon^{i j}_{\phantom{i j} k} \widehat{\mu_g}\left( {S^k_1}\right) .
	\label{2_120}
\end{equation}
A complete description of quantization in LQG, including the quantization of a volume, can be found, for instance, in the textbooks~\cite{Rovelli:2014ssa,Gambini:2011zz};
we will not go into this subject here.

\section{Quantization of measure in the general case
}
\label{quant_gen}

In this section we suggest a method of quantization of measure in the general case, analogous to what was done in LQG when quantizing an area.
The main idea is as follows:
\begin{itemize}
\item the triangulation of spacetime is introduced, as in LQG;
\item the measure $\mu$ is introduced, which in general is not associated with metric;
\item the operator of this measure $\hat{\mu}$ is introduced, for which the commutation relations are imposed,
analogous to what was done for the operator of area in LQG.
\item we use the Radon-Nikodym theorem, which relates two measures
$d \mu$ and $d \mu_g$.
\end{itemize}
Thus we have the measure $\mu$ related to the measure $\mu_{g}$ associated with the metric $g$ as follows:
$$	
	\mu(\mathcal A) = \int\limits_{\mathcal A} f d \mu_g .
$$
This can be rewritten in the infinitesimal form as
$$	
	d \mu = f d \mu_g ,
$$
where $f$ is some function, called the Radon-Nikodym derivative. After the quantization, we arrive at the following operator equality:
$$	
	\widehat {\mu(\mathcal A)}	= \int\limits_{\mathcal A} \hat{f} \, \widehat{d \mu_g} .
$$
Formally, in the infinitesimal form, this can be rewritten as
$$	\widehat{d \mu} = \hat{f} \, \widehat{d \mu_g}  .$$
The restriction of this measure to the area labelled by the link $\ell_i$ has the following form:
$$	\widehat{\mu(S^i)} = \int \limits_{S^i} \widehat{d \mu^i} = \int \limits_{S^i}\hat{f} \widehat{d \mu_g^i} .$$
Now we are dealing with the very difficult question of defining the properties of the operator $\widehat{\mu(S^i)}$. The challenge we face is the presence
of the operator $\hat{f}$; this does not allow us to generalize the  commutation relations~\eqref{2_120} in a simple way.

The main problem here is the determination of the properties of the operator $\hat{f}$. Of course, in the general case this operator can be an arbitrary quantity,
but it is only of interest to consider the case where the operator $\hat{f}$ is defined using some equations. In this case we have only one possibility:
$\hat{f}$ is a function of some quantized field whose properties can be defined using the corresponding quantization rules for the given field.
For example, this can be a positively defined Lagrangian of such a field,
$\hat{f} = \mathcal L(\hat \Phi^A)$, where the index $A$ corresponds to all possible indices of the field $\Phi^A$.
The positiveness of the Lagrangian is imposed due to the theorem~\ref{Radon_Nikodym_1}.
As such positively defined Lagrangian, one can take, for example, a Lagrangian in Euclidean space with the field potential bounded below.

As another possibility, one can consider an approach with the so-called dynamical measure introduced in Refs.~\cite{Guendelman:1996qy,Guendelman:1996jr,Guendelman:2012gg}.
In these works, it is assumed that, in the gravitational action,  instead of the measure  $d \mu_g$ associated with metric, one uses a nonmetric measure
$
	d \phi^1 \wedge d \phi^2 \wedge d \phi^3 \wedge d \phi^4
$ with the scalar fields $\phi^i, i = 1,2,3,4$, which are determined using some dynamical equations.

As one more possibility, one can consider modified bimetric gravity \cite{Hassan:2011zd} where two measures corresponding to the metrics $g_1$ and $g_2$
are used  in the gravitational action to integrate the Ricci scalars $R\left( g_2\right)$ and $R\left( g_1\right)$, respectively, as follows:
$$	\mathcal L \propto \sqrt{- g_1 } R\left( g_2\right) + \sqrt{- g_2 } R\left( g_1\right).$$
This enables one to use in each term the measure which is not associated with metric in the corresponding scalar curvature.
Nevertheless, for each measure, there are its own dynamical differential Euler-Lagrange equations.
Within this approach, as the $\Phi^A$-fields for the metric $g_1$, it will be the metric $g_2$ and vice versa.

In the aforementioned case when the operator $\hat{f}$ is taken to be $\hat{f} = \mathcal L(\hat \Phi^A)$, the algebra of operators of the measure
$\widehat{\mu(S)}$ becomes much more complicated compared with the algebra of operators of the measure
 $ \widehat{\mu_g(S)}$ associated with metric, since it is necessary to take into account the algebra of operators of the quantized field $\hat{\Phi}^A$.
 In the general case the properties of the operator  $\widehat{\mu}$ will be determined by all Green functions
 \begin{equation}
	G_n = \left\langle \Psi_{\hat{\mu}} \right|
		\hat{\mu}\left( S_1\right), \hat{\mu}\left( S_2\right), \dots , \hat{\mu}\left( S_n\right)
	\left| \Psi_{\hat{\mu}} \right\rangle ,
\label{3_80}
\end{equation}
where $	\left| \Psi_{\hat{\mu}} \right\rangle $ is a quantum state describing both the measure $\widehat{\mu_g}$
and the quantized field $\hat{\Phi^A}$. It is important to note here that the properties of the operator
$\widehat{\mu(S)}$ are not determined by the commutation relations similar to~\eqref{2_120}, but should be determined by all Green functions~\eqref{3_80}.

The situation is simplified considerably if the operator $\hat{f}$ is just a function of $\hat{f} = f$.
In this case one can write the following equality that generalizes the corresponding theorem from the mathematical analysis:
$$	\widehat{\mu(S)} = \int \limits_{S} f \, \widehat{d \mu_g}
	= f(\xi)  \int \limits_{S} \widehat{d \mu_g} , \quad \xi \in S .$$
Then, using the commutation relations~\eqref{2_120}, one can derive the following expression:
\begin{equation}
\begin{split}
	\left[
		\widehat{\mu(S^i_1)} , \widehat{\mu(S^j_2)}
	\right] = & i \Delta_{S_1 S_2} l_0^2
	\frac{f\left( \xi_1\right) f\left( \xi_2\right) }{f\left( \xi_3\right) }
	\epsilon^{i j}_{\phantom{i j} k} \widehat{\mu}\left( {S^k_1}\right)
	=  i \Delta_{S_1 S_2} l_0^2
	F\left( \xi_{1,2,3} \right)
	\epsilon^{i j}_{\phantom{i j} k} \widehat{\mu}\left( {S^k_1}\right)
\\
		= & i \Delta_{S_1 S_2} l^2 \left( \xi_{1,2,3}\right)
	\epsilon^{i j}_{\phantom{i j} k} \widehat{\mu}\left( {S^k_1}\right) ,
\end{split}\nonumber
	\label{3_100}
\end{equation}
where $ \xi_1 \in S^i_1$, $ \xi_2 \in S^j_2$, and $ \xi_3 \in S^k_1$,
and a nonlocal quantity $l \left( \xi_{1,2,3}\right)$ with the dimensions
of length occurs. Formally, in the infinitesimal form, this expression takes the form
$$	\left[
	\widehat{d \mu^i} , \widehat{d \mu^j}
	\right] = i \Delta_{dS_1 dS_2} l_0^2
	f (x) \epsilon^{i j}_{\phantom{i j} k} \widehat{d \mu^k}
	=  i \Delta_{dS_1 dS_2} l^2 (x) \epsilon^{i j}_{\phantom{i j} k} \widehat{d \mu^k} ,$$
since in the infinitesimal form $\xi_1 = \xi_2 = \xi_3 = x \in dS_1$ and one can introduce the function $l(x)$ with the dimensions of length.

The appearance of the function  $l \left( \xi_{1,2,3}\right)$ instead of the fundamental length $l_0$ has the result that the discrete spectrum~\eqref{2_90}
contains now the length $l$ which varies  in space; this leads to interesting consequences. For example, near the singularity,
$
	f\left(x \rightarrow x_{\text{sing}} \right) \rightarrow \infty
$, and hence the length
$
	l\left(\xi_{1,2,3} \rightarrow x_{\text{sing}} \right) \rightarrow \infty
$
becomes infinitely large as well. That is, the size of an elementary cell in the spin foam becomes infinitely large. But since the singularity occurs at the point
and the size of the elementary cell becomes infinitely large while approaching the singularity (at that, one cannot speak about the space inside the cell),
this means that while approaching the singularity an increase of the function $f$ is regulated such that it remains finite inside the cell.
This means that the singularity is regularized when the quantum measure is taken into account.

Using the same methods as those employed in LQG (see, for example,  Ref.~\cite{Rovelli:2014ssa}, chapters 5 and 8),
one can quantize the measure $\widehat{d \mu}$ restricted to the volume and length.

\section{Physical theories on the background of space with quantum measure
}

LQG suggests that our space is a quantum object with fluctuating measure, and, as we clarified here, the measure need not be associated with metric.
Therefore, the question arises as to how one can define the action in classical theory or quantize fields when a quantum measure is taken into account?

The next interesting question is whether it is possible to make somehow a particular choice of the function  $f$ in the classical case or of
the operator $\hat{f}$ in the quantum case? If we compare this situation with classical physics, we see that there are fundamental equations
coming from the principle of least action which determine the fields, and in quantum theory there are quantization rules determining
the properties of field operators. Thus we are interested in the following question: how one can assign equations determining the function $f$ or the operator $\hat{f}$?

We think that the only way of doing this is as follows: in the classical case, $f$ is a function of some fields, i.e., it is a Lagrangian density;
in the quantum case, $\hat{f}$ is defined by operators of some quantum field whose algebra must be defined independently.
Then in the classical case the Euler-Lagrange equations determine the fields $\rightarrow$ the Lagrangian  $\rightarrow$ the function $f$ $\rightarrow$ the operator of the measure
 $f \widehat{d \mu_g}$. In the quantum case, one defines the algebra of field operators  $\rightarrow$ the operator $f$ $\rightarrow$ the operator of the measure $\hat{f} \widehat{d \mu_g}$.

\subsection{Classical physics on the background of spacetime with quantum measure
}

Consider now the question of how one can assign equations of motion in classical physics on the background of spacetime with quantum measure.
One may assume that this can be done as follows. The quantum states in LQG are described by the spin network states
$
	\psi_{ j_\ell, v_n} (\mathcal{U}_\ell) = \left| \Gamma , j_\ell, v_n
	\right\rangle ,
$
where $\Gamma$ is the graph, $v$ are the eigenvalues, $\left|  v \right\rangle $ are eigenstates of the volume in each Hilbert space $\mathcal K_{j_1 \ldots j_4}$
(for more details concerning these quantities, see Ref.~\cite{Rovelli:2014ssa}, chapter 7). In each such state, one can write an action for any classical field and then,
using the principle of least action, calculate the field distribution ensuring the minimum of the corresponding action.
In practice, this means that one obtains a solution to the Euler-Lagrange equations for the given classical field. Thus we have the following solution to the classical Euler-Lagrange equations:
\begin{equation}
		\frac{\partial}{\partial x^\mu} \left(
			\frac{\partial \mathcal L}{\partial \Phi^A_{, \mu}}
		\right) - \frac{\partial \mathcal L}{\partial \Phi^A_{\mu}} = 0
		\text{ on the background of the spin network state } \left| \Gamma , j_\ell, v_n	\right\rangle ,
\label{4_a_10}
\end{equation}
where the index $A$ in the field $\Phi^A$ is a generalized index labelling all indices of the given field. As a result, we have a solution
of the Euler-Lagrange equations~\eqref{4_a_10} $\phi^A_{\left| \Gamma , j_\ell, v_n	\right\rangle }$ on the background of the spin network state
$\left| \Gamma , j_\ell, v_n	\right\rangle$. Physically, this means that for a pure quantum state $\left| \Gamma , j_\ell, v_n	\right\rangle$
the spatial distribution of the classical field  $\Phi^A$ is $\phi^A_{\left| \Gamma , j_\ell, v_n	\right\rangle }$.
If we have a quantum superposition
$$	\Psi = \sum_{\left| \Gamma , j_\ell, v_n	\right\rangle } C_{j_\ell, v_n}
	\left| \Gamma , j_\ell, v_n	\right\rangle ,$$
the mean field $\Phi^A$ may be determined in a standard manner  to be
$$	\left\langle \Phi^A \right\rangle = \sum_{\left| \Gamma , j_\ell, v_n	\right\rangle }
	\left| C_{j_\ell, v_n} \right|^2 \Phi^A_{\left| \Gamma , j_\ell, v_n	\right\rangle }.$$
Similarly, one can determine the dispersion
$
	\left\langle \left(
			\left\langle \Phi^A \right\rangle - \Phi^A_{\left| \Gamma , j_\ell, v_n	\right\rangle }
	\right)^2
	\right\rangle
$ and all other average values.

\subsection{Quantum field theory on the background of space with quantum measure
}

In this subsection we consider the question of how one can define quantum field theory on the background of space with quantum measure.
This means that we have a space endowed with quantum measure which is associated with metric and given by the commutation relations
in the infinitesimal form~\eqref{2_110} or in the finite form~\eqref{2_120}, and also the quantized field $\hat{\Phi^A}$.

In this case a wave function describing such a system will be the product of the wave function describing the quantum state of the field
$\Phi^A$,  $\left| \Psi_{\Phi^A} \right\rangle $, and  the spin network state,
$
\psi_{ j_\ell, v_n} (\mathcal{U}_\ell) = \left| \Gamma , j_\ell, v_n
\right\rangle
$,
$$	\left| \Psi_{\hat{\mu}, \Phi^A} \right\rangle =
	\left| \Gamma , j_\ell, v_n \right\rangle
	\left|  \Psi_{\Phi^A} \right\rangle .$$
The presence of this wave function enables us, analogous to what was done in Sec.~\ref{quant_gen}, to define the Green function
\begin{equation}
	G_n \left(
		x_1, x_2, \dots , x_n; S_1, S_2, \dots , S_n
	\right)
	= \left\langle \Psi_{\hat{\mu}} \right|
		\Phi^A\left( x_1\right) \widehat{\mu_g}\left( S_1\right),
		\Phi^A\left( x_2\right) \widehat{\mu_g}\left( S_2\right),
		\dots ,
		\Phi^A\left( x_n\right) \widehat{\mu_g}\left( S_n\right)
	\left| \Psi_{\hat{\mu}} \right\rangle .
\label{4_b_20}
\end{equation}
As one can see, the structure of these Green functions is extremely complicated since they contain both the local arguments $x_i$ and nonlocal ones:
the restriction of the measure $\widehat{\mu_g}$ to two-dimensional areas $S_i$.

But this is not all. Using the techniques developed in LQG, it is necessary to define the restriction of the operator of the measure
 $\widehat{\mu_g}$ to the one- and three-dimensional regions (volumes and lengths). Then, using these operators and making a generalization,
 as in Eq.~\eqref{4_b_20}, one can define the corresponding Green functions
$$	G_n \left(
	x_1, x_2, \dots , x_n; O_1, O_2, \dots , O_n
	\right)
	= \left\langle \Psi_{\hat{\mu}} \right|
	\Phi^A\left( x_1\right) \widehat{\mu_g}\left( O_1\right),
	\Phi^A\left( x_2\right) \widehat{\mu_g}\left( O_2\right),
	\dots ,
	\Phi^A\left( x_n\right) \widehat{\mu_g}\left( O_n\right)
	\left| \Psi_{\hat{\mu}} \right\rangle ,$$
where $O_i$ are one-, two-, and three-dimensional regions. These Green functions will completely determine the properties of the operator
 $\widehat{\mu} = \hat{f} \widehat{\mu_g}$.

\section{Physical discussion and conclusions}

One of remarkable results obtained in LQG is the quantization of length, area, and volume; this implies the possibility of quantization of measure
which is associated with metric. Using the Radon-Nikodym theorem relating any two measures, we have shown that in gravitation
one can quantize any measure which is not even associated with metric. In this case the properties of the operator of a measure which is not associated with metric
differ from the properties of the operator of a measure associated with metric. In the simplest case, if the Radon-Nikodym derivative
(the proportionality coefficient between two measures in the Radon-Nikodym theorem) is some function, the fundamental length appearing in the right-hand side of
the corresponding commutation relations becomes a variable quantity. For example, near a singularity where
the Radon-Nikodym derivative goes to infinity, the size of an elementary cell in the spin foam will tend to infinity, and this can lead to smoothing the singularity.
In the case where the Radon-Nikodym derivative is an operator, the properties of the operator of a measure change  even
more stronger, and they are determined by all Green functions of the operator of a measure.

For such an approach in quantizing the measure, the following interesting question arises. We have shown that, based on the results obtained within LQG,
one may quantize any measure which can be even not associated with metric. The quantization of the measure which is not associated with metric
is carried out through Green functions~\eqref{3_80} using the quantization of the measure associated with metric in the form~\eqref{2_120}
or in the infinitesimal form~\eqref{2_110}. Here, it is of great importance that \textit{the measure is a nonlocal object and hence nonlocal objects are present in the commutation relations}.
The regions where the nonlocal objects are defined can intersect one another or not, and also they can have different dimensions.
All this greatly complicates the problem of defining the commutation relations for such nonlocal objects.
To solve this extremely complicated problem, in LQG, in order to make the problem as simple as possible, one introduces a triangulation of space.
In practice, the triangulation enables one to assign commutation relations only for regions having identical dimensions.
Therefore, a natural question arises as to whether the triangulation is an auxiliary trick introduced only to quantize nonlocal objects?
If it is true, the space remains continuous (i.e., not discrete), and the triangulation is just a convenient trick, allowing one to avoid all problems related to the quantization of nonlocal objects.
This resembles somewhat the situation in general relativity where one can use any coordinate systems, but there are invariant quantities, e.g., the scalar curvature.

Next, we have considered
classical and quantum theories on the background of space endowed with quantum measure.
In the first case, we have shown that for any quantum state of the spin foam there is its own solution of the Euler-Lagrange field equations.
This means that in the general case there is a quantum  superposition of solutions of the Euler-Lagrange  equations, for which one can evaluate the average solution,
its dispersion, etc. In the second case, the given quantum system is determined by quantum states of the spin foam and quantized field.

The question of quantization of measure is the question of quantization of a nonlocal quantity.
The literature in the field discusses the question of nonlocal theories for a long time (see, e.g., Refs.~\cite{Efimov:1967pjn,Moffat:1990jj,Modesto:2017sdr}),
and therefore, in the context of the present paper, the question arises as to whether there exists  any relation between the quantization of measure and nonlocal quantum field theories?
This interesting question requires a special and careful consideration.

A special and separate question arises when one studies the quantization of metric and measure. In LQG, it is assumed  that the approach being suggested there
will enable one to quantize gravity as well, that is, a metric and connection. If this is not so, then the question arises concerning the quantization of two independent quantities~--
the measure and the metric. The main challenge appears when one takes into account their mutual influence on each other, when both quantities are quantum objects.
In the classical case this problem is resolved in such a way that the relation $d \mu_g = \sqrt{-g} d^4 x$ between them is postulated.
An additional difficulty is that, in the classical case, there are no Euler-Lagrange  equations for the measure which
give the relation between the measure and metric  $d \mu_g = \sqrt{-g} d^4 x$, similar to the relation between the metric and connection within the Palatini approach.

In conclusion, we would like to note that in the present paper we have discussed positively defined measures, but, using the theorem~\ref{Radon_Nikodym_2},
one can generalize the results obtained by us to signed and complex measures.

\section*{Acknowledgments}

The work was supported by the program No.~AP14869140  (The study of QCD effects in non-QCD theories) of the Ministry of Education and Science of the Republic of Kazakhstan. We are also grateful to the Research Group Linkage Programme of the Alexander von Humboldt Foundation for the support of this research.

\appendix

\section{Minimum necessary insight into measure theory
}	
Definition of measure:
\begin{definition}
	Let $X$ be a set and $\Sigma$ a $\sigma$-algebra over $X$. A function $\mu$ from $\Sigma$ to the extended real number line is called a measure if it satisfies the following properties:
	\begin{itemize}
		\item Non-negativity: For all $E$ in $\Sigma$, we have $\mu(E) \geq 0$.
		\item Null empty set: $ \mu (\varnothing )=0$.
		\item Countable additivity (or $\sigma$-additivity): For all countable collections
		$\left\lbrace E_{k}\right\rbrace_{k=1}^{\infty }$ of pairwise disjoint sets in $\Sigma$,
$$			\mu \left(
				\bigcup_{k=1}^{\infty } E_{k}
			\right) = \sum_{k = 1}^{\infty } \mu (E_{k}) .$$
	\end{itemize}
\end{definition}
The pair $\left( X, \Sigma\right) $ is called a measurable space. A triple $\left( X, \Sigma, \mu\right) $ is called a measure space.
The Radon-Nikodym theorem involves a measurable space $(X,\Sigma )$ on which two $\sigma$-finite measures are defined, $\mu$ and $\nu$.
\begin{theorem}[the Radon-Nikodym theorem~\cite{Rudin}]
	 if $\nu \ll \mu$ (that is, if $\nu$  is absolutely continuous with respect to $\mu$), then there exists a $\Sigma$-measurable function $f:X\to [0,\infty )$,
such that for any measurable set $\mathcal A \subseteq X$,
	 \begin{equation}
	 		\nu (\mathcal A) = \int\limits_{\mathcal A} f\,d\mu .
	 \end{equation}
\label{Radon_Nikodym_1}
\end{theorem}
The function $f$ is called the Radon-Nikodym derivative and is denoted as $\frac{d \nu}{d \mu}$.
The meaning of this theorem is that any two measure are interrelated by some function.


A similar theorem can be proven for signed and complex measures:
\begin{theorem}[the Radon-Nikodym theorem]
	if $\mu$ is a nonnegative $\sigma$-finite measure, and $\nu$ is a finite-valued signed or complex measure such that $\nu \ll \mu$, that is, $\nu$
is absolutely continuous with respect to $\mu$, then there is a $\mu$-integrable real- or complex-valued function $g$ on $X$ such that for every measurable set $\mathcal A \subseteq X$
	\begin{equation}
		\nu (\mathcal A) = \int\limits_{\mathcal A} g\,d\mu .
	\end{equation}
\label{Radon_Nikodym_2}
\end{theorem}

\end{document}